\documentclass[preprint,5p,number]{elsarticle} 

\usepackage{graphics}
\usepackage{amsmath}
\usepackage{float}
\usepackage{subfig}
\usepackage{tabularx}
\usepackage[normalem]{ulem}

\begin{document}
\title{The effect of pump depletion on reversible photodegradation.}
\author{Benjamin Anderson, Sheng-Ting Hung and  Mark G. Kuzyk}
\address{Department of Physics and Astronomy, Washington State University,
Pullman, WA 99164-2814}
\date{\today}

\begin{abstract}
We model the effect of pump depletion on reversible photodegradation using the extended domain model[Anderson and Kuzyk, arXiv:1309.5176v1, \textbf{2013}] and the Beer-Lambert law. We find that neglecting pump absorption in the analysis of the linear optical transmittance leads to an underestimate of the degree and rate of photodegradation. The model is used to accurately measure the molecular absorbance cross sections of the three species involved in photodegradation of disperse orange 11 dye in (poly)methyl-methacralate polymer (DO11/PMMA). Finally we find that the processing history of a dye-doped polymer affects reversible photodegradation, with polymerized monomer solutions of DO11 being more photostable than those prepared from solvent evaporated dye-polymer solutions.

\vspace{1em}

\end{abstract}

\maketitle

\vspace{1em}

\section{Introduction}
Reversible photodegradation of dye-doped polymers is a relatively new phenomena reported in dye polymer combinations of disperse orange 11 (DO11) doped into PMMA \cite{howel02.01,howel04.01,embaye08.01,Ramini11.01,Ramini11.02,Ramini12.01, Ramini13.01,Anderson11.01,
Anderson12.01,Anderson12.02,Anderson13.01,Anderson13.03}, DO11 doped into polystyrene\cite{Hung12.01}, various anthraquinone derivatives doped into PMMA \cite{Anderson11.02}, AF455 doped into PMMA\cite{Zhu07.01,Desau09.01}, pyrro-methene and rhodamine doped PMMA\cite{peng98.01}, and 8-hydroxy-quinoline aluminum (Alq3) doped into PMMA\cite{Kobrin04.01}.  Quantitative measurements of reversible photodegradation assume that samples are sufficiently thin for the pump intensity to be constant throughout the sample \cite{Zhu07.01,embaye08.01, Anderson11.01,Anderson11.02,Ramini12.01,Ramini13.01,Anderson13.03}.  In this study we use the extended correlated chromophore domain model (eCCDM)\cite{Anderson13.03} and the Beer-Lambert law to account for pump depletion.  Additionally, we use this model to find the absorbance cross sections of the three species involved in reversible photodegradation.

\section{Theory}
The effect of pump absorption on {\em irreversible} photodegradation has been studied extensively with the degradation mechanism assumed to be the conversion of undamaged molecules into a different molecular species\cite{Vydra96.01,Simmons71.01,Zhang98.01,Rezzonico07.01,Rezzonico08.01,Kim94.01,Moshrefzadeh93.01}.  For a monochromatic pump beam, the degradation process is described by a simple rate equation

\begin{align}
\frac{dn}{dt}&=-\frac{c\epsilon_0(\omega_p)}{\hbar \omega_p B(\omega_p)}nI_p,
\\ &=-\alpha n I_p, \label{eqn1}
\end{align}
where $n$ is the fractional population of undamaged molecules, $c$ is the concentration of absorbers, $\hbar \omega_p$ is the pump photon energy, $B$ is frequency-dependent bleaching rate\cite{Gonzalez00.01,Gonzalez00.02,Gonzalez01.01,Gonzalez03.01}, $\epsilon$ is the molecular absorptivity of the undamaged species, $\alpha=\frac{c\epsilon_0}{\hbar \omega_p B}$ is the intensity-independent decay rate, and $I_p$ is the pump intensity.  The pump intensity as a function of depth follows the differential Beer-Lambert law,

\begin{equation}
\frac{dI_p}{dz}=-n(t)\sigma_0I_p-[1-n(t)]\sigma_1I_p,  \label{eqn2}
\end{equation}
where $\sigma_0$ is the absorbance per unit length of the undamaged species and $\sigma_1$ is the absorbance per unit length of the damaged species.

Equations \ref{eqn1} and \ref{eqn2} have been used to model photodegradation in three cases: 1) neglecting pump absorption ($\sigma_0=\sigma_1=0$) \cite{Gonzalez00.02,Gonzalez99.01}, 2) assuming a transparent damaged species ($\sigma_1=0$)\cite{Vydra96.01,Simmons71.01,Dubois96.01} and assuming an absorbing damaged species \cite{Zhang98.01,Rezzonico07.01,Rezzonico08.01,Kim94.01,Moshrefzadeh93.01}. Cases 1 and 2 are found to have analytic solutions, while case 3 requires numerical solution methods.

In this study we use the three-species eCCDM model \cite{Anderson13.03} with the assumption that all three species absorb pump light.  The eCCDM proposes that domains of correlated chromophores are responsible for self-healing after photodegradation.  These domains are found to be described by a linear aggregation model with the distribution of domains size $N$ being $\Omega(N)$\cite{Ramini12.01,Ramini13.01,Anderson13.03,raminithesis,andersonthesis}.  For a domain of size $N$, the population dynamics at depth $z$ are modeled by a three-species ``parallel'' degradation model, with rate equations given by\cite{Anderson13.03}

\begin{align}
\frac{\partial n_0(z,t;N)}{\partial t}=&-\left(\frac{\alpha}{N}+\epsilon N\right) I_p(z,t) n_0(z,t;N) \nonumber
\\ &+\beta N n_1(z,t;N), \label{eqn:n0domD}
\\ \frac{\partial n_1(z,t;N)}{\partial t}=&\frac{\alpha I_p(z,t)}{N}n_0(z,t;N)-\beta N n_1(z,t;N),
\\ \frac{\partial n_2(z,t;N)}{\partial t}=&\epsilon N I_p(z,t) n_0(z,t;N), \label{eqn:n2dom}
\end{align}
where $n_0(z,t;N)$ is the undamaged population, $n_1(z,t;N)$ is the reversibly damaged population, $n_2(z,t;N)$ is the irreversibly damaged population, $\alpha$ is the intensity independent reversible decay rate, $\epsilon$ is the intensity independent irreversible decay rate, $I_p(z,t)$ is the pump intensity at depth $z$, and $\beta$ is the recovery rate.  Equations \ref{eqn:n0domD} through \ref{eqn:n2dom} describe the microscopic dynamics, while the macroscopic dynamics are determined by an ensemble average over all domains

\begin{equation}
\overline{n}_i=\sum_{N=1}^\infty n_i(z,t;N)\Omega(N),
\end{equation}
where $\Omega(N)$ is the density of domains of size $N$.

To describe the effect of pump absorption on decay and recovery we us the  differential Beer-Lambert law to describe the pump and probe intensity depth profiles:

\begin{align}
 \frac{\partial I_p(z,t)}{\partial z}&= -I_p(z,t)\sum_{N=1}^{\infty}\bigg[n_0(z,t;N)\sigma_0(\omega_p) \nonumber
\\ & +n_1(z,t;N)\sigma_1(\omega_p) +n_2(z,t;N)\sigma_2(\omega_p)\bigg]\Omega(N),
\\ \frac{\partial I(z,t;\omega)}{\partial z}&= -I(z,t;\omega)\sum_{N=1}^{\infty}\bigg[n_0(z,t;N)\sigma_0(\omega) \nonumber
\\ &+n_1(z,t;N)\sigma_1(\omega)+n_2(z,t;N)\sigma_2(\omega)\bigg]\Omega(N),  \label{eqn:domID}
\end{align}
where $\sigma_i(\omega)$ is the absorbance per unit length at frequency $\omega$ of the $i^{th}$ species with $i=0$ being the undamaged species, $i=1$ the reversibly damaged species, $i=2$ the irreversibly damaged species, and $\omega_p$ is the pump frequency.  

\section{Experimental Method}
We measure the effect of pump depletion on photodegradation using several thin films with differing thickness of (poly)methyl-methacralte (PMMA) doped with disperse orange 11 (DO11) to a concentration of 9g/l.  Several different preparation methods are used to produce a wide range of thickness.  Thickness measurements are performed using absorbance spectroscopy, and transmittance imaging microscopy is used to measure photodegradation.

\subsection{Samples prepared from monomer}
The thickest samples are prepared using bulk dye-doped polymer as follows.  Filtered methyl-methacralate (MMA)  is mixed with DO11 dye in the correct proportions to obtain 9g/l.  The solution is then sonicated for half an hour, at which point initiator (butanethiol) and a chain transfer agent (Tert-butyl peroxide) are added in amounts of 33$\mu$l per 10ml of MMA, and the solution is sonicated for another 30-60 min.  After sonication the solution is filtered through 0.2 $\mu$m disk filters to remove particulates, and put into glass vials.  The vials of solution are placed in a 95$^\circ$C oven for 48 hours to complete polymerization. 

Small pieces of the bulk dye-doped polymer are removed and thermally pressed between two glass substrates to form a thick film.  A custom oven/sample press applies an uniaxial stress of 90 psi at a temperature of 150$^\circ$C for one hour, allowing the polymer melt to uniformly flow from the center, at which point the stress is gradually removed while the sample is allowed to cool.  Typical thicknesses for this preparation method ranges from  60-100 $\mu$m.

\subsection{Samples prepared from polymer solution}
To form thinner samples we add dye to PMMA/solvent solution, which is then either spun coated or thermally pressed after a drop is placed on a substrate.  The dye-doped PMMA/solvent solution is prepared as follows.  DO11 and PMMA in a ratio of 9g/l are dissolved into a solution of 33\% $\gamma$-butyrolactone and 67\% propylene glycol methyl ether acetate (PGMEA) with a ratio of 15\% solids to 85\% solvents.  The solution is stirred for 72 hr to dissolve the dye and polymer, after which the solution is filtered with 0.2$\mu$m disk filters into vials to remove any remaining solids.  

1.5 cm $\times$ 1.5 cm glass substrates are flooded with solution and then spun at 1200 rpm for 30s /layer, with thicker films requiring multiple layers.  After spin coating, the samples are placed in an 85$^\circ$C oven for 24 hours to force solvent evaporation and remove other volatilities, after which they are allowed to cool.  Spin coating typically results in films between 1-10 $\mu$m.

To get a thicknesses between 10 $\mu$m and 60 $\mu$m we use the method of drop pressing.  Drop pressing involves heating a glass substrate at 50$^\circ$C for 10 mins, at which point  DO11/PMMA/solvent solution is dropped onto the substrate, and the temperature is raised to 95$^\circ$C for half an hour to induce solvent evaporation.  The sample is then placed in a vacuum oven at room temperature overnight to ensure the sample is dry.  Once dried, the sample is used to make a sandwich structure with another clean glass substrate and placed in the thermal press oven for 135 mins at a uniaxial stress of 72psi and a temperature of 130$^\circ$C.

\subsection{Thickness Measurements}
Sample thickness is measured using the pristine absorbance, $A$, of each sample, which is related to the thickness, $L$, by:

\begin{equation}
A=\epsilon_0 c L
\end{equation}
where $c$ is the concentration and $\epsilon_0$ is the molecular absorbance cross section of undamaged DO11/PMMA which is  extrapolated from the absorption spectrum of DO11/MMA \cite{raminithesis,andersonthesis}. The absorbance measurements are performed at six different locations on the samples to determine the average thickness, with the  average thicknesses for the four samples used in these studies of $8\pm 2$ $\mu$m, $22\pm 3$ $\mu$m, $35\pm 5$ $\mu$m, and $83\pm 8$ $\mu$m.

\subsection{Degradation Measurements}
A CW Ar:Kr laser, operating at 488 nm, focused to a line with a peak intensity of 120 W/cm$^2$ induces degradation, which is probed using a blue LED and transmittance imaging microscopy\cite{Anderson11.01,Anderson11.02}.   The transmittance imaging microscope measures the change in absorbance due to photodegradation, which we denote as the scaled damaged population, $n'(t)$.  The scaled damaged population (SDP), in terms of probe intensity, $I$, is:

\begin{equation}
n'(t)=-\ln \left[\frac{I(L,t)-I(L,0)}{I(L,0)}\right],  \label{eqn:sdp}
\end{equation}
where $I(L,0)$ is the initial intensity transmitted before degradation, and $I(L,t)$ is the intensity transmitted at time $t$ during degradation.

\subsection{Absorbance cross section measurements}
Photodegradation and recovery is measured at different pump intensities on several 9g/l DO11 samples prepared using the polymerization method.  Photodegradation is induced with a Verdi Nd:Yag CW laser, whose beam is spatially filtered to produce a circular TEM$_{00}$ Gaussian beam of 2mm diameter.  The probe beam is an Ocean Optics PX-2 Pulsed Xenon source focused to a 0.75mm diameter spot centered on the pump beam.  Spectra during decay and recovery are measured using an Ocean Optics SD2000 spectrometer.

For two damaged species, the change in absorbance, $\Delta A(t;\omega)$, during photodegradation and recovery may be written as

\begin{equation}
\Delta A(t;\omega)=f\left[\Delta \sigma_1\int_0^L\overline{n}_1 dz+\Delta \sigma_2 \int_0^L\overline{n}_2dz\right]  \label{eqn:absdiff}
\end{equation}
where $\Delta\sigma_i=\sigma_i-\sigma_0$ and $f$ is the pump-probe correlation factor\cite{andersonthesis}, which represents the fraction of probe beam area that is uniformly damaged.  For our experimental configuration $f=0.9$.

\section{Results and discussion}

\subsection{Fluence Independent Recovery Rate}
\begin{figure}
\centering
\includegraphics{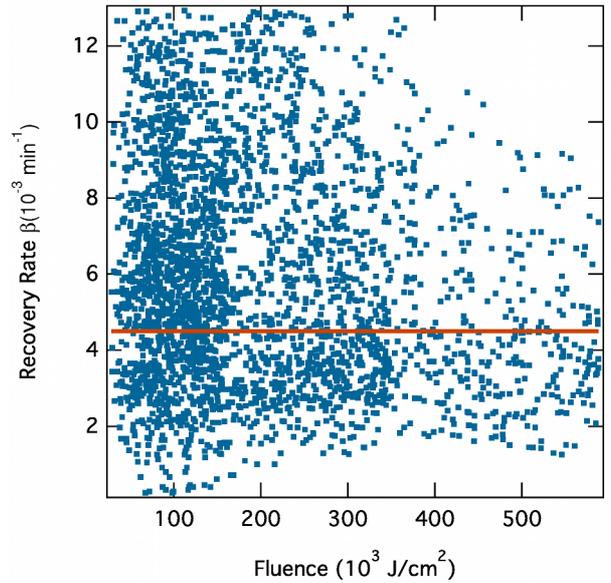}
\caption{Experimental single exponential recovery rate as a function of fluence (points) and theory (line).}
\label{fig:flurec}
\end{figure}

\begin{figure}
\centering
\includegraphics{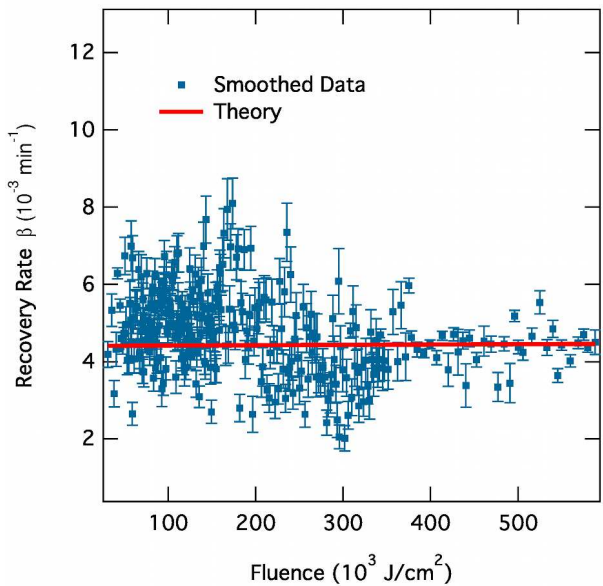}
\caption{Recovery rates smoothed using weighted averaging over twenty adjacent points.}
\label{fig:smoothed}
\end{figure}

Previously we showed that the simple exponential recovery rate is dose independent for a limited range of pump fluence\cite{Anderson11.01,Anderson11.02}.  To  test the dependence of the recovery rate over a larger range of pump fluence, we measure the recovery rate of a 9g/l DO11/PMMA sample prepared from monomer using fluences of $\sim$ 25-600 kJ/cm$^2$.  Figure \ref{fig:flurec} shows the raw simple exponential recovery rate as a function of fluence measured at 5000 points across the sample, with Figure \ref{fig:smoothed} showing the recovery rates smoothed using twenty point weighted averages.  Both Figures \ref{fig:flurec} and \ref{fig:smoothed} show a linear fit with slope within experimental uncertainty of zero.  The large degree of scatter is due to point-to-point fluctuations within the sample.

\begin{figure}
\centering
\includegraphics{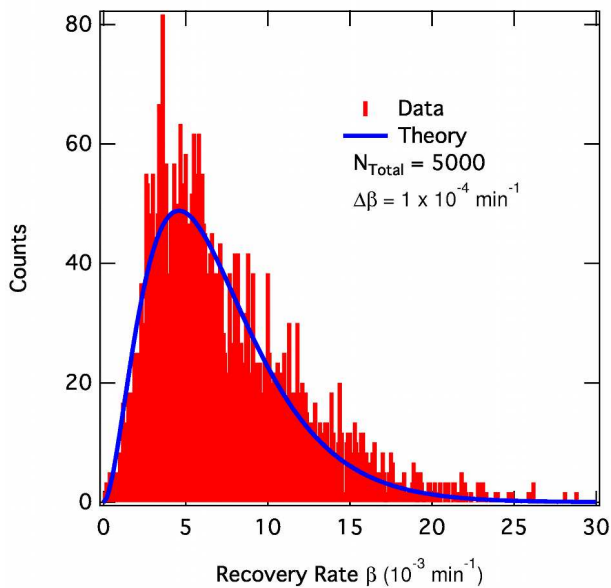}
\caption{Histogram of recovery rates measured at 5000 points over a 9g/l DO11/PMMA sample and fit to a poisson distrbution}
\label{fig:hist}
\end{figure}

Figure \ref{fig:hist} shows the distribution of recovery rates using a histogram with a bin width of $\Delta \beta = 1 \times 10^{-4}$min$^{-1}$ with a poisson fit as a guide for the eye.  The weighted average over all points gives a recovery rate of $\overline{\beta}= 4.502(\pm 0.045)\times10^{-3}$ mins$^{-1}$, which is within experimental uncertainty of recovery measurements with low fluence ($\sim$0.1 kJ /cm$^2$), where $\beta = 4.0( \pm 1.5)\times10^{-3}$ mins$^{-1}$ \cite{Ramini12.01,raminithesis,Ramini13.01}.

\subsection{Thickness Dependent Photodegradation}
\begin{figure}
\centering
\includegraphics{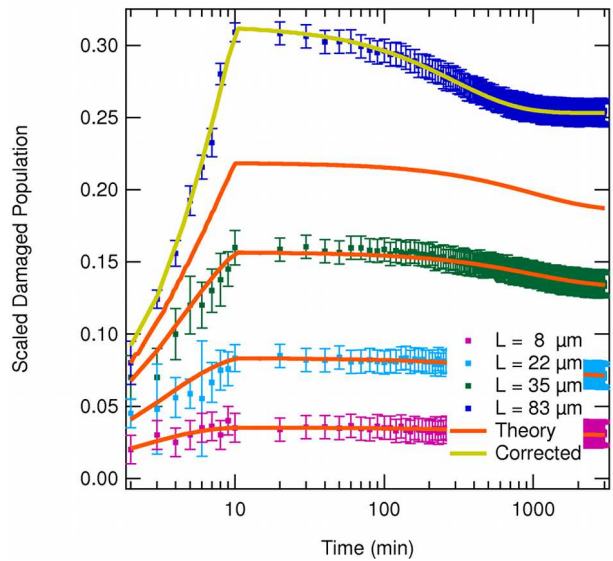}
\caption{Scaled damaged population as a function of time (points) and theory (curves) using parameters determined from the 22$\mu$m sample. The curve ``corrected'' uses a different parameter set to fit the thickest sample.}
\label{fig:sdp}
\end{figure}

SDP decay and recovery curves are measured at ten different fluences for each of the four samples and fit to Equation \ref{eqn:sdp} by using Equations \ref{eqn:n0domD} through \ref{eqn:domID}  to determine the transmitted probe intensity as a function of time.  The data for the 8$\mu$m, 22$\mu$m and 35$\mu$m samples are found to fit the model with one set of parameters, with the thickness the only parameter that depends on sample, shown as points and solid orange curves in Figure \ref{fig:sdp}. However, the 83$\mu$m sample is found to require a different parameter set from the other three samples (most notably the recovery rate is three times as fast).  A fit using a parameter set different from the other three thicknesses is shown as a yellow curve in Figure \ref{fig:sdp}.   Table \ref{Tab:depthfit} shows the fit parameters for the 8$\mu$m, 22$\mu$m and 35$\mu$m samples, as well as the fit parameters for the 83$\mu$m sample, and the three-species model parameters  -- excluding the effects of pump absorption -- for the 22 $\mu$m sample for comparison.

To explain the discrepancy in the models parameters we note that the 8$\mu$m, 22$\mu$m and 35$\mu$m samples are made using the solution method, while the 83$\mu$m sample is made using the polymerization method. We hypothesize that the method of preparation changes the decay and recovery characteristics.   The sample prepared by polymerization is found to be more photostable because the polymer and dye may be more strongly interacting when the polymer polymerizes around the dye.  In contrast, the dyes in a co-dissolved solution are left behind in the region defined by pockets of solvent.

\begin{table*}
\begin{center}
\begin{tabular}{|c|c|c|c|cl|}
\hline 
\multicolumn{6}{|c|}{\textbf{Model Parameters}}\\ \hline
& Solution& Monomer & No Depth\cite{Anderson13.03} & & \\[1em] \hline
$\alpha$ & 15.5  ($\pm$ 3.9) & 11.2  ($\pm$ 1.1) & 1.32  ($\pm$ 0.33)  & $\times$ & 10$^{-2}$ cm$^2$/W min  \\[1em]
$\beta$  &  2.28 ($\pm$ 0.95)  &6.5 ($\pm$ 1.3) & 2.53 ($\pm$ 0.51)    &$\times$ & 10$^{-5}$ min$^{-1}$ \\[1em]
$\epsilon$  &  2.06 ($\pm$ 0.61)  &1.29  ($\pm$ 0.25) &   0.647 ($\pm$ 0.021)    &$\times$ & 10$^{-5}$ cm$^2$/W min  \\[1em] \hline
$\sigma_0$ &  \multicolumn{2}{|c|}{56.00 ($\pm$ 0.05)} & -  & $\times$ & $10^{-3}$ $\mu$m$^{-1}$ \\[1em]
$\sigma_1$ & \multicolumn{2}{|c|}{54.5  ($\pm$ 1.5)}& - & $\times$  & $10^{-3}$ $\mu$m$^{-1}$\\[1em]
$\sigma_2$ & \multicolumn{2}{|c|}{21.3  ($\pm$ 5.5)}&  - & $\times$  & $10^{-3}$ $\mu$m$^{-1}$\\[1em]
 \hline
\end{tabular}
\end{center}
\caption{Depth effect model parameters for solvent prepared samples using $\lambda=0.29$eV and $\rho=0.019$.  Parameters for the 22$\mu$m sample (ignoring the depth effect) are included for comparison.}
\label{Tab:depthfit}
\end{table*}

Aside from the differences in samples prepared from monomer and polymer, we also note that the decay rate parameters found taking the effect of pump absorption into account are larger than those neglecting the effect of pump absorption.  This result is expected as pump absorption will result in lower photodegradation rates deeper in a sample where the pump intensity is lower.  Additionally, the recovery rate is found to be independent of sample thickness, which is consistent with the observation that the recovery rate is independent of absorbed fluence.

\begin{figure}
\centering
\includegraphics{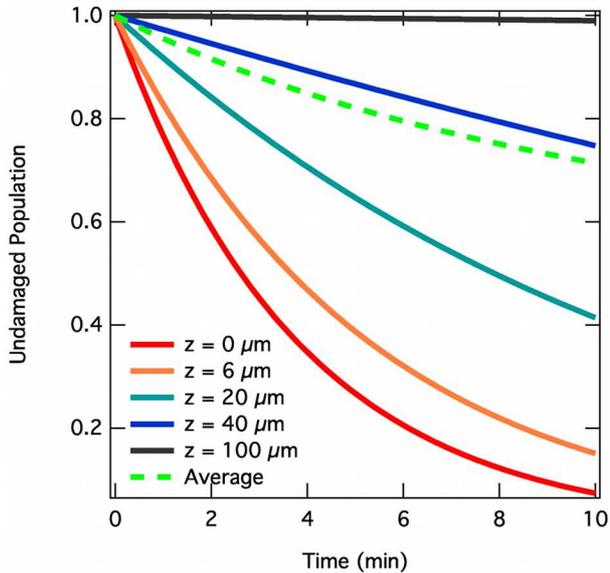}
\caption{Calculated undamaged population as a function to time for several depths (solid curves) and the average population (dashed curve) which would be obtained from optical measurements.}
\label{fig:pop}
\end{figure}

So far we have considered the scaled damaged population, which represents an average over the entire thickness of the sample.  To better understand the actual population dynamics at various depths in the sample, we solve Equations \ref{eqn:n0domD} and \ref{eqn:n2dom} using the experimentally determined parameters for the samples prepared from the polymer/dye solution.  Figure \ref{fig:pop} shows a simulation of the undamaged population decay as a function of time for five different depths, for total thickness of 100$\mu$m and 120 W/cm$^2$ intensity.  As the depth increases both the degree and rate of decay decreases.  This change in population dynamics, as a function of depth, leads to the average population underestimating the true decay rate and degree of damage.

\subsection{Molecular Absorbance Cross Sections}
\begin{figure}
\centering
\includegraphics{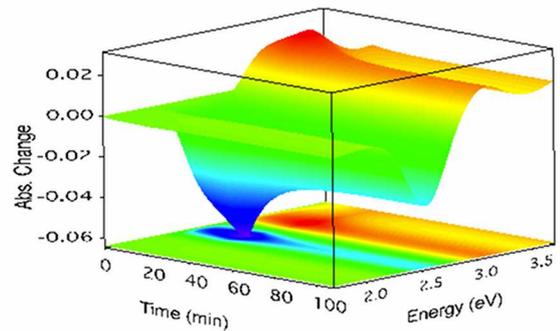}
\caption{Experimental fit to typical change in absorbance as a function of time and probe energy.}
\label{fig:absdiff}
\end{figure}

\begin{figure}
\centering
\includegraphics{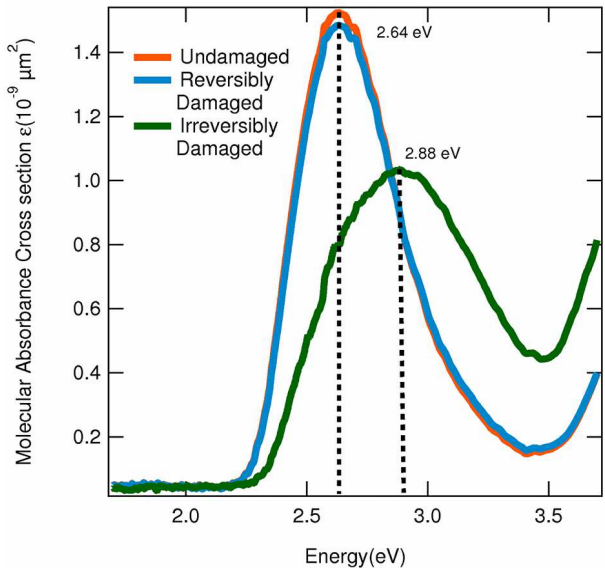}
\caption{Molecular absorbance cross sections of the three observed species as a function of energy.}
\label{fig:3lcross}
\end{figure}

We fit the measured change in absorbance during decay and recovery to Equation \ref{eqn:absdiff} for several sample thicknesses and intensities, with an example fit shown in Figure \ref{fig:absdiff}. Using the fit parameters for each thickness and intensitiy we calculate the average molecular absorbance cross sections of the damaged species, as shown in Figure \ref{fig:3lcross}.  The undamaged cross section is included for comparison.

Figure \ref{fig:3lcross} shows that the reversibly damaged species has a similar cross section spectrum to the undamaged species, and the irreversibly damaged species has a drastically different absorbance cross section.  This is consistent with cross-sections determined from imaging measurements, shown in Table \ref{Tab:depthfit}.  There are two important features of the irreversibly damaged species cross section: (1) the visible peak is blue shifted by 0.24eV, and (2) the absorbance in the UV regime is larger.  The observation of a blue-shifted peak is consistent with the hypothesis that charged molecular fragments are formed\cite{Desau09.01,Anderson13.01,Anderson13.03}, and the increased absorbance in the deep-blue/UV is consistent with the proposal that the irreversible damaged species are damaged polymer chains\cite{Anderson13.03} as it is well known that photodegraded neat PMMA has an increase in absorbance in the deep-blue/UV regime\cite{Baum07.01,Baum10.01,Ahmed09.01} as well as increased scattering\cite{Watanabe12.01}, which is greater at higher energies.

\section{Conclusions}
Using the eCCDM and the differential Beer-Lambert law we model the effect of pump absorption on reversible photodegradation.  We find that neglect of pump depletion leads to an underestimation of the decay rate and degree of decay.  However, the rate of recovery is found to be thickness invariant, as it is fluence independent.

We have also measured the absorbance cross sections of the damaged species in  DO11/PMMA, taking both pump absorption and pump probe overlap into account. The reversibly damaged species' cross section is found to be similar to the undamaged species' cross section, and the irreversibly damaged species' absorbance peak is found to be blue shifted, as well as having a greater absorbance cross section in the near UV.

Finally, we find that the method of sample preparation (i.e. from solution or from monomer) changes the decay and recovery characteristics, with samples prepared from monomer being more robust.  We propose that polymerizing monomer in the presence of dyes results in better domain formation, as the monomer may polymerize around the dye molecules.  However, in the case of PMMA and dye in solution, domain formation is less complete, as the PMMA chains do not form around the dye but rather, the dyes are left in place when solvent evaporates.  Additionally, incomplete solvent evaporation may contribute to these observations.  Experiments are currently underway to explore the effects of varying dye, polymer and processing techniques in order to better understand domain formation.

\section{Acknowledgments}
We thank Wright Patterson Air Force Base and Air Force Office of Scientific Research (FA9550- 10-1-0286) for their continued support of this research.

\newpage

\bibliographystyle{model1-num-names}
\bibliography{PrimaryDatabase}

\end{document}